\documentclass[twocolumn]{aastex62}
\turnoffedit

\usepackage{empheq,cases}

\accepted{2023 December 24}
\submitjournal{AAS journals}

\shorttitle{Chondrule survivability in the solar nebula}
\shortauthors{Taki \& Wakita}

\begin{document}

\title{Chondrule survivability in the solar nebula}

\correspondingauthor{Tetsuo Taki}
\email{takitetsuo@gmail.com}

\author[0000-0002-6602-7113]{Tetsuo Taki}
\affiliation{School of Arts \& Sciences, University of Tokyo, 3-8-1, Komaba, Meguro, 153-8902 Tokyo, Japan}
\affiliation{Center for Computational Astrophysics, National Astronomical Observatory of Japan, Osawa, Mitaka, Tokyo 181-8588, Japan}

\author[0000-0002-6602-7113]{Shigeru Wakita}
\affiliation{Department of Earth, Atmospheric and Planetary Sciences, Massachusetts Institute of Technology, Cambridge, MA, 02139, USA}



\begin{abstract}
The lifetime of mm size dust grains, such as chondrules, in the nominal solar nebula model is limited to $\sim 10^{5}$ yr due to an inward drift driven by gas drag.
However, isotopic and petrological studies on primitive meteorites indicate a discrepancy of $\gtrsim 10^{6}$ yr between the formation time of chondrules and that of chondritic parent bodies.
Therefore chondrules should survive for $\gtrsim 10^{6}$ yr in the solar nebula against the inward drift without subsequent growth (i.e., planetesimal formation).
Here we investigate the conditions of the solar nebula that are suitable for the long lifetime of chondrule-sized dust particles.
We take the turbulent strength, the radial pressure gradient force, and the disk metallicity of the solar nebula as free parameters.
For 1 mm-radius-chondrules to survive and keep their size for $\gtrsim 10^{6}$ yr, the suitable condition is a weak turbulence ($\alpha \sim 10^{-6}$), a flat radial profile ($\eta \lesssim 10^{-3}$), and a high metallicity ($Z\sim 0.1$).
This condition is qualitatively consistent with the characteristics of protoplanetary disks suggested by recent observations.
We eventually propose that planetesimal formation may be induced by the disk evolution, e.g., the inside-out dispersal of the gas component due to the disk wind.
\end{abstract}



\section{Introduction} \label{sec:intro}
We call building blocks of planets as planetesimals.
Although planetesimal formation processes are still unclear
\edit2{
\citep[e.g.,][]{2014prpl.conf..547J,2020RAA....20..164L,2023ASPC..534..717D},
}
the planet formation is established by assuming the existence of planetesimals \citep[e.g., $N$-body simulations;][]{2000Icar..143...15K}:
Planetesimals which are on the order of kilometers, collide with each other and grow up into protoplanets as the runaway growth.
Then, protoplanets which are on the order of a few thousand kilometers, become a planet through the oligarchic growth.

When materials originating from planetesimals are preserved throughout the history of the solar system, they provide a key to understanding the history of solar system formation.
Previous works indicated that asteroids are collections of remnants of planetesimals that survived from further growth and destructive events during their lives \citep{2002Icar..156..399B,Kobayashi2016a}.
Thus, we could say asteroids record the history of the solar system.
The sample return missions from asteroids would be the best for exploring the early solar system, such as Hayabusa, Hayabusa 2, and OSIRIS-REx.
In fact, the Itokawa's sample given by the Hayabusa mission revealed that they resemble primitive meteorites \citep[\edit1{ordinary chondrites,}][]{2011Sci...333.1113N,2015aste.book..397Y}.
\edit1{
Additionally, the components of Ryugu's sample brought back by Hayabusa 2 mission are similar to other primitive meteorites \citep[carbonaceous chondrites,][]{2022NatAs...6..214Y,doi:10.1126/science.abn7850}.
}
While the returned sample is limited, we can analyze a large number of meteorites.
Since they keep the fossil record of the solar system, they are also important objects to understand the early age of the solar system.
In particular, primitive meteorites are indirect evidence for the existence of planetesimals in the past solar system.

We group meteorites by
\edit1{
their chemistry, isotopic compositions, and texture
}
\citep[e.g.,][]{2014mcp..book...65S}.
Among them, chondrites are one of the primitive meteorites group.
Chondrites contain early formed unique grains, Ca-Al-rich inclusions (CAIs) and chondrules \citep[e.g.,][]{dac14}.
The isotopic analyses showed that the CAIs condensed at the beginning of the solar system, which is 4567 million years ago \citep[e.g.,][]{2010E&PSL.300..343A,cbk12}.
Chondrules formed at from a just after the formation of CAIs to 5 million years after CAIs  \citep[e.g.,][]{cbk12,bbk15}.
These grains formed in the solar nebula, then accreted on the parent bodies of meteorites, planetesimals.

\begin{figure}[ht]
\begin{center}
\includegraphics[width=8cm]{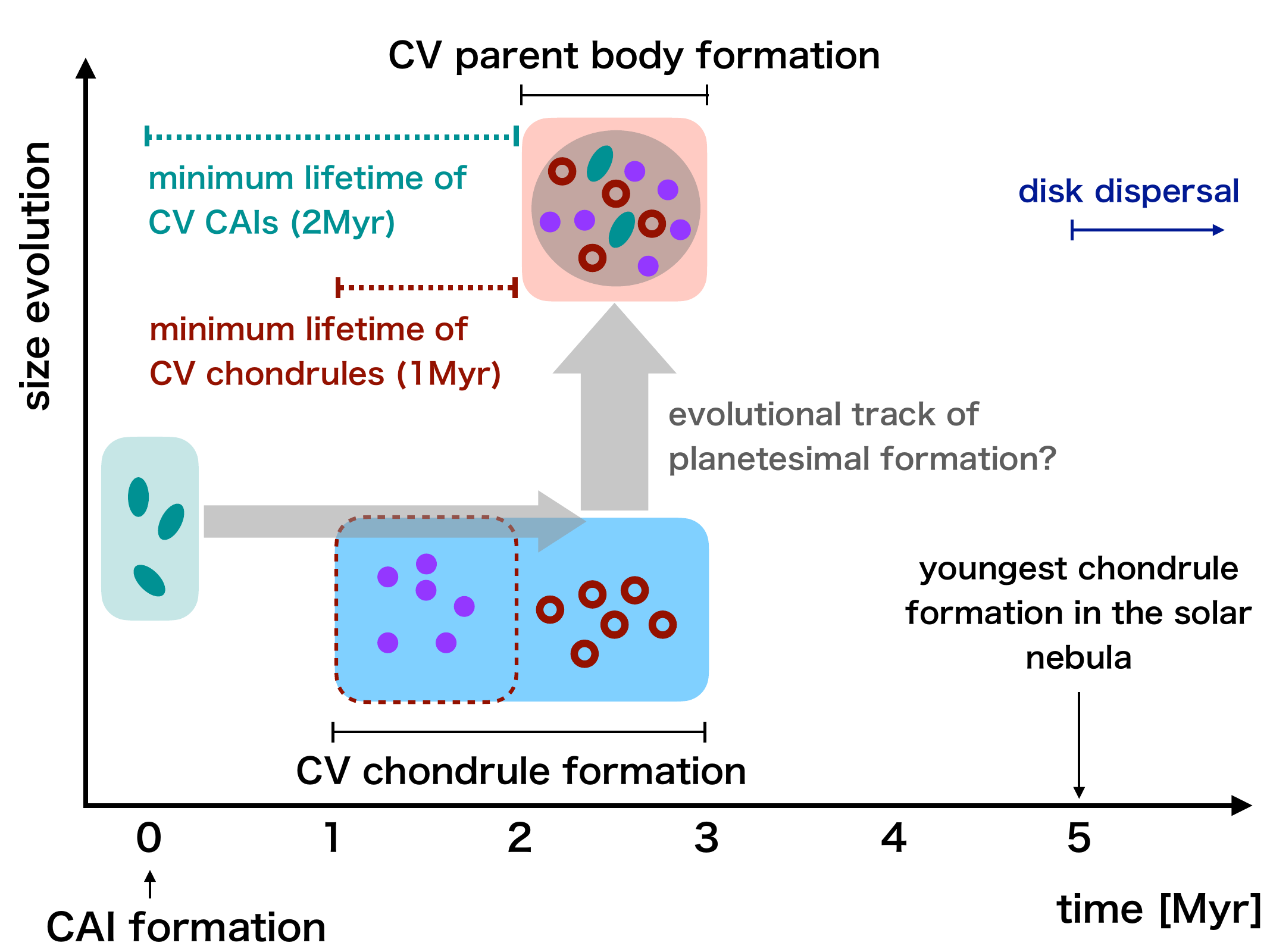}
 \caption{
Schematic illustration of the timeline of chondrule and parent body formation in the solar nebula.
The time gap between the formation of small particles and that of the planetesimals which contain such small particles indicates the long lifetime of small particles.
This figure also depicts the timing of gas dissipation in the solar nebula suggested by the youngest (CB) chondrules (see footnote \ref{fnlabel}).
}
 \label{fig:timeline}
\end{center}
\end{figure}

Previous studies suggest the chondritic parent bodies formed at a few million years after CAIs,
combining the isotopic studies and thermal modeling \citep[e.g.,][]{dac14,2014prpl.conf..571G}.
For example, isotopic ages of secondary minerals and internal heating model by decay heat of $^{26}$Al suggested that parent bodies of CV chondrites accreted 2-3 million years after the CV CAIs
\edit1{
\citep{2014M&PS...49..772S,djn15,jni17}.
}
Since CV chondrules formed at 1-3 million years after the formation of CAIs \citep{cbk12}, chondrules can form contemporaneous with the parent bodies of chondrites \citep[e.g.,][]{2019GeCoA.260..133T,2017GeCoA.201..303N}.
Even \edit1{if} this happens, there is a time gap between the formation of CV CAIs and their host parent bodies.
Additionally, the variation of chondrule ages within a single chondrite is also reported which is about 1 million years \citep{cbk12,2009Sci...325..985V}.
Given these evidences, CAIs and chondrules should survive 1 million years at least in the solar nebula before accreting onto the parent bodies 
\edit1{
\citep[e.g.,][]{2012Icar..220..162J,2015MNRAS.452.4054G}.
}
Figure~\ref{fig:timeline} illustrates the timeline of CV parent body formation as discussed above.

CAIs and chondrules are large grains as the first generation grains in the early solar nebula.
Typical size of CAIs and chondrules ranges from $0.1 \ {\rm mm}$ to $1.0 \ {\rm cm}$ \citep{2014mcp..book..139M,Krot:2009aa,Friedrich:2015aa,Simon:2018aa}.
Although their formation process is still under debate \citep[e.g.,][]{2018ApJS..238...11D}, we can discuss their size evolution processes as solid particles\footnote{We here use the term ``solid particles'' to refer to both the smallest unit of the solid component (i.e., silicate monomers and chondrules) and the aggregates they comprise.}.
The maximum size of solid particles is basically controlled by two effects; the fragmentation (or bouncing) at the collision and the radial drift due to the gas drag.
When we consider the direct sticking (i.e., mutual collisions) as a growth mechanism for solid particles, too fast collisions prevent them from further growth because such high-speed collisions induce the fragmentation \citep{1993Icar..106..151B} or bouncing \citep{2010A&A...513A..57Z} of solid particles rather than their growth.
The fragmentation (or bouncing) indirectly supports the discrepancy between the formation age of chondrules and that of parent bodies of chondrites.
A relative velocity between small particles in the protoplanetary disk becomes larger as they grow.
The relative velocity of 1 mm sized particles is roughly comparable to the critical fragmentation velocity of silicate dust aggregates \citep[e.g.,][see also Section~\ref{sec:relat-veloc-betw}]{Birnstiel2012} in the minimum mass solar nebula \citep[MMSN;][]{Hayashi1981a}.
Therefore, the growth of CAIs and chondrules due to the direct sticking is difficult once they are formed.
They should keep their original sizes until some additional growth mechanisms (e.g., self-gravitational collapse) set in.

A radial drift of small dust particles, however, inhibits them to stay in the solar nebula.
The circumstellar disks, including the solar nebula, usually have a negative pressure gradient in a radial direction.
The gas component rotates as a sub-Keplerian flow because of the dynamical equilibrium.
On the other hand, dust particles tend to rotate at the Keplerian velocity.
Since dust particles feel the disk gas as a headwind due to their different velocities, they lose their angular momentum.
As a result, dust particles move toward their host star and cannot stay in the protoplanetary disk
\edit1{
\citep{Adachi1976a, Weidenschilling1977a,2012A&A...537A..61L}.
}
A typical timescale of the radial drift for $\sim$ 1 mm-size grain is 0.1 million years, which is shorter than the survival time for chondrules of 1 million years (see Section~\ref{sec:radi-drift-timesc} for more details).

\edit1{
The survival time of small particles, including chondrules, in the protoplanetary disk has been studied from various perspectives.
\citet{2007A&A...469.1169B,2008A&A...480..859B} showed that a high dust-to-gas mass ratio can solve the issue of radial migration of small particles.
\citet{2012Icar..220..162J} and \citet{2015MNRAS.452.4054G} particularly focused on the gap in the formation ages between chondrules and chondrite parent bodies in the solar nebula.
They pointed out that the effect of significant radial diffusion makes the chondrules' evolution consistent with meteoritic evidence.
However, it should be included that high dust-to-gas mass ratios can trigger more rapid planetesimal formation than expected (See Section~\ref{sec:stre-inst}).
In addition, strong radial diffusion may be inconsistent with a recent picture of disk evolution derived from the non-ideal magnetohydrodynamic calculations \citep[e.g.,][]{Bai2017a}.
}

\edit1{
The goal of this study is to search for properties of solar nebulae consistent with chondrule lifetimes.
We employ a simple strategy that slows down the radial migration of chondrules.
To slow down the chondrules' migration, we change the parameters of the solar nebula to the extent that inhibits planetesimal formation.
}
We introduce our models to examine the lifetime of chondrules in Section~\ref{sec:model}.
Section~\ref{sec:result} shows the favorable disk parameters for chondrules.
In Section~\ref{sec:discussion}, we discuss the limitations of our model and compare the disk parameters we have obtained with recent radio observations of protoplanetary disks.
Implications for the formation of planetesimals are also discussed in Section~\ref{sec:discussion}.
Our conclusions are presented in Section~\ref{sec:conclusions}.

\section{Model}
\label{sec:model}

\subsection{Disk model}
\label{sec:disk-model}

Since the lifetime of the solar nebula is much longer than that of chondrules, we can treat the solar nebula as a steady gas disk for considering the chondrule evolution.
The youngest chondrules found in CB chondrites\footnote{These chondrules are not the same as those in the CV chondrites and are not the subject of survivability studies in this paper. We address the CB chondrules as the evidence which suggests that the disk gas was present 5 million years after the formation of the CV CAIs (Fig. \ref{fig:timeline}). \label{fnlabel}} are formed in the gaseous disk about $5 \times 10^{6}$ yr after the formation of CAIs \citep{bbk15,2005Natur.436..989K,2017SciA....3E0407B}.
The lifetime of typical protoplanetary disks would be $\sim 10^{7}$ yr based on the observed disk frequency \citep[e.g.,][]{2011ARA&A..49...67W}.
Recent studies proposed a formation age of Jupiter, which requires a large amount of gas for \edit1{its} formation, is about $2-3 \times 10^{6}$ yr after CAIs \citep{2017ApJ...841L..17G,2017PNAS..114.6712K}.
These facts suggest that the lifetime of the solar nebula would be much longer than a few million years after CAIs.
We now focus on the survival time of chondrules in CV chondrites, $t_{\rm surv} \sim 10^6$ yr, which is in the first few million years of the solar system (see Section~\ref{sec:surv-cond-small}).
Therefore, our assumption of the steady solar nebula is reasonable in considering the survivability of CV chondrules.

We adopt the MMSN model as a reference of the solar nebula \citep{Hayashi1981a}.
In this model, the disk temperature $T$ is
\begin{eqnarray}
 T \approx 280 \left(\frac{r}{1\, {\rm au}}\right)^{-1/2} \left( \frac{L}{L_{\sun}}\right)^{1/4} {\rm K},
\label{eq:temp_mmsn}
\end{eqnarray}
where $r$ is the orbital radius, $L$ is the luminosity of the central star, and $L_{\sun}$ is the luminosity of \edit1{the} current Sun, respectively.
We assume $L = L_{\sun}$ for simplicity.
The gas surface density is given by
\begin{eqnarray}
 \Sigma_{{\rm g, MMSN}} = 1.7 \times 10^3 \left(\frac{r}{1 \, {\rm au}}\right)^{-3/2} {\rm g \ cm}^{-2}.
\label{eq:sigma_gas_mmsn}
\end{eqnarray}
Assuming the hydrostatic balance in a vertical direction and locally isothermal gas,
\edit1{the gas scale height, $h_{\rm g}$,}
and the gas pressure at the disk midplane, $P$, are given by
\edit1{
\begin{eqnarray}
 h_{\rm g} = c_{\rm s} / \Omega \approx 0.033 \left( \frac{r}{1 \, {\rm au}} \right)^{5/4} \, {\rm au},
\end{eqnarray}
and,}
\begin{eqnarray}
 P = c_{\rm s}^2 \rho_{{\rm g},0} \approx 1.4 \times 10^{1} \left( \frac{r}{1 \, {\rm au}} \right)^{-13/4} \ {\rm g \ cm}^{-1}{\rm s}^{-2},
\end{eqnarray}
where $c_{\rm s}$ is the sound speed of gas, \edit1{where $\Omega$ is the Keplerian frequency,} and $\rho_{{\rm g},0}$ is the spatial density of gas at the disk midplane, respectively.

Since the disk gas rotates as a sub-Keplerian flow due to the radial pressure gradient, the azimuthal velocity of gas can be given by $v_{\phi} = (1-\eta)v_{\rm K}$, where $v_{\rm K}$ is the Keplerian velocity, and $\eta$ is a dimensionless pressure gradient force in the radial direction \citep{Adachi1976a,Nakagawa1986a},
\begin{eqnarray}
 \eta \equiv \frac{-\partial P/\partial r}{2\Omega v_{\rm K} \rho_{{\rm g},0}}.
\label{eq:eta_def}
\end{eqnarray}
\edit1{
Note that the dimensionless pressure gradient force is a fundamental and widely used parameter considering the radial velocity of dust \citep[e.g.,][]{Adachi1976a,2014prpl.conf..339T}.
}
In the MMSN model, $\eta$ takes a small value as
\begin{eqnarray}
 \eta_{\rm MMSN} \approx 1.8 \times 10^{-3}.
\label{eq:eta_mmsn}
\end{eqnarray}
Since we consider an inner region of the solar nebula (a few au), we can ignore a dependence of $\eta_{\rm MMSN}$ on $r$.
We discuss the validity of this assumption in Section~\ref{sec:radial-location-cv}.

\subsection{Radial drift timescale of dust particles}
\label{sec:radi-drift-timesc}

\edit2{
The radial drift velocity of dust particles is derived as
\begin{eqnarray}
 v_{{\rm d},r} = - \frac{2\tau_{\rm stop}}{(1+\epsilon)^2 + \tau_{\rm stop}^2} \eta v_{\rm K}
                 + \frac{1+\epsilon}{(1+\epsilon)^2 + \tau_{\rm stop}^2} u_{\nu},
  \label{eq:NSH}
\end{eqnarray}
where $\epsilon$ is the dust-to-gas mass ratio at the disk midplane, $\tau_{\rm stop} \equiv t_{\rm stop}\Omega$ is the dimensionless stopping time of dust particles normalized by $\Omega^{-1}$, and $u_{\nu}$ is the gas radial velocity due to the viscous accretion, respectively \citep{2016A&A...591A..72I}.
}
We now call $\tau_{\rm stop}$ as the Stokes number.
The stopping time of dust particles, $t_{\rm stop}$, is expressed as \citep[e.g.,][]{Sato2016a},
\begin{eqnarray}
 t_{\rm stop} =   \frac{\rho_{\rm int}a}{\rho_{{\rm g},0} v_{\rm th}} \max \left(1, \frac{4a}{9\lambda_{\rm mfp}} \right),
\label{eq:t_stop_def}
\end{eqnarray}
where $\rho_{\rm int}$, $a$, $v_{\rm th}$ and $\lambda_{\rm mfp}$ are the internal density of dust particles, the radius of dust particles, the thermal velocity of the disk gas, and the mean free path of the gas molecules, respectively.
We assume $\rho_{\rm int}=3.0 \ {\rm g \ cm}^{-3}$ which is the nominal value for chondrules \citep{2015ChEG...75..419F}.
We use $v_{\rm th} = \sqrt{8 k_B T / (\pi m_g)}$, where $k_B$ is the Boltzmann constant and $m_g = 3.9 \times 10^{-24} {\rm g}$ is the mean molecular mass of disk gas.
We take $\lambda_{\rm mfp} = m_g / (\sigma_{\rm mol}\rho_{{\rm g},0})$, where $\sigma_{\rm mol}=2 \times 10^{-15}{\rm cm}^2$ is the molecular collisional cross section.
When $a < 9\lambda_{\rm mfp}/4$, the gas drag is in the Epstein regime, and otherwise if $a > 9\lambda_{\rm mfp}/4$, it is in the Stokes regime.
The Stokes number in the MMSN model can be given as
\begin{eqnarray}
 \tau_{\rm stop} = 2.71 && \times 10^{-4} \left( \frac{r}{1\, {\rm au}} \right)^{3/2}
 \left( \frac{a}{1\, {\rm mm}} \right)  \nonumber \\
 && \max \left[1, \ \frac{a}{3.13 \left(r/1 \, {\rm au} \right)^{11/4}{\rm cm}} \right].
\label{eq:stokes}
\end{eqnarray}

We define the radial drift timescale of dust particles as
\edit2{
\begin{eqnarray}
&& t_{\rm drift} \equiv \frac{r}{\mid v_{{\rm d},r} \mid} \nonumber \\
&& \sim \min \left[\frac{(1+\epsilon)^2 + \tau_{\rm stop}^2}{2\tau_{\rm stop}}\frac{r}{\mid \eta v_{K} \mid}, \  
   \frac{(1+\epsilon)^2 + \tau_{\rm stop}^2}{1+\epsilon}\frac{r}{\mid u_{\nu} \mid} \right ]. \nonumber \\
\,
  \label{eq:drift_time}
\end{eqnarray}
}
Typical size of CAIs and chondrules ranges from $0.1 \ {\rm mm}$ to $1.0 \ {\rm mm}$ \citep{2014mcp..book..139M,Krot:2009aa,Friedrich:2015aa,Simon:2018aa}.
Even though a few CAIs have $\sim 1.0 \ {\rm cm}$ \citep{2014mcp..book..139M,Hezel:2008aa}, we can safely employ the Epstein drag regime for these particles when they locate in $r \gtrsim 0.7$ au.
Since their Stokes number is much lower than unity, $\tau_{\rm stop} \ll 1$, we can drop the term of $\tau_{\rm stop}^2$ in Equation (\ref{eq:drift_time}).
\edit2{
In addition, the effect of gas accretion is almost negligible in this work (see Appendix~\ref{appendixA}).
We can simply assume that the first element inside the brackets in Equation~(\ref{eq:drift_time}) is always chosen.
}

Then, we obtain the approximate expression of $t_{\rm drift}$:
\begin{eqnarray}
 t_{\rm drift} \approx && 1.63 \times 10^5 \nonumber \\
&& (1+\epsilon)^2 \left(\frac{ \eta }{\eta_{\rm MMSN}}\right)^{-1}
\left(\frac{a}{1 \, {\rm mm}}\right)^{-1}{\rm yr}.
    \label{eq:fiducial_drift_time}
\end{eqnarray}
Here $t_{\rm drift}$ does not depend on $r$, but it does if we consider the $r$-dependency of $\eta$ (see Section~\ref{sec:radial-location-cv}).
Assuming nominal values that $\epsilon \ll 1, \ \eta = \eta_{\rm MMSN}$, and $\ a=1 \ {\rm mm}$, $t_{\rm drift}$ takes about $10^5$ yr, which is the referred value in Section~\ref{sec:intro}.
To explain the time gap between the formation age of the small particles (i.e., CAIs and chondrules) and that of parent bodies of meteorites, those small particles should stay for at least $10^6$ yr in the solar nebula.
As a result, the estimated $t_{\rm drift}$ is far shorter than the survival time of small particles, $t_{\rm surv} = 10^6$ \ yr.
\edit1{
Equation (\ref{eq:fiducial_drift_time}) indicates that the chondrule lifetime should be increased by about one order of magnitude above the nominal value.
To achieve this lifetime extension, we need to combine two effects (i.e., $\epsilon$ and $\eta$).
Because the impact of changing only one of them is limited, it is not enough to extend the lifetime of chondrules (see Section~\ref{sec:stre-inst} and \ref{sec:plaus-disk-struct}).
}

\subsection{Dust-to-gas mass ratio}
\label{sec:dust-gas-mass}

To obtain $\epsilon$, we first determine the vertical profile of the dust density.
Dust particles tend to settle on the disk midplane owing to the vertical gravity from the central star.
When the turbulence exists in the solar nebula as classically considered in protoplanetary disks \citep[e.g.,][]{2011ARA&A..49..195A}, the turbulence stirs up the dust particles as diffusion.
The balance between the sedimentation and diffusion of dust particles determines the dust scale height.
Since the dust particles sufficiently couple with the gas to move at terminal velocity, both the degree of sedimentation and diffusion depends on the Stokes number.
Hence, the dust scale height is a function of the turbulent strength and the Stokes number of dust particles.

\citet{Youdin2007b} derived the dust scale height, $h_{\rm d}$, which is determined by the balance between the sedimentation and diffusion, including the effects of orbital and vertical oscillations:
\begin{eqnarray}
 h_{\rm d} \approx h_{\rm g} \left( 1+ \frac{\tau_{\rm stop}}{\alpha}\frac{1+2\tau_{\rm stop}}{1+\tau_{\rm stop}}\right)^{-1/2},
\end{eqnarray}
where
$\alpha$ is the dimensionless diffusion parameter in a vertical direction.
We assume that $\alpha$ is the same as $\alpha$-parameter for a viscous evolution of an accretion disk \citep{Shakura1973a}.

When we adopt $1$ mm for the typical size of chondrules, their Stokes number is $\tau_{\rm stop} \sim 10^{-4} \ll 1$ at $1$ au.
Oscillations are ineffective for such the small particles because of $(1+2\tau_{\rm stop})/(1+\tau_{\rm stop})\approx 1$.
Therefore we can use the simple form of $h_{\rm d}$:
\begin{eqnarray}
 h_{\rm d} \approx h_{\rm g} \left( 1+ \frac{\tau_{\rm stop}}{\alpha}\right)^{-1/2}.
  \label{eq:h_particle_app}
\end{eqnarray}
When $\tau_{\rm stop} \ll \alpha$, which corresponds to cases with the strong turbulence of gas and/or cases with the strong coupling between the particle and gas motion, the dust scale height agrees with the gas scale height, $h_{\rm d} \approx h_{\rm g}$.
Instead, $h_{\rm d}$ becomes small and finally converges to zero in the case with $\tau_{\rm stop} \gg \alpha$.

The disk metallicity, $Z$, can be defined as $Z \equiv \Sigma_{\rm d}/\Sigma_{\rm g}$, where $\Sigma_{\rm d}$ and $\Sigma_{\rm g}$ is the dust and gas surface density, respectively.
When we assume that the spatial density of gas ($\rho_{\rm g}$) and that of dust ($\rho_{\rm d}$) have the Gaussian profile in the vertical direction, i.e., $\rho_{\rm d}=\Sigma_{\rm d}/\left(\sqrt{\pi}h_{\rm d}\right)$ and $\rho_{\rm g}=\Sigma_{\rm g}/\left(\sqrt{\pi}h_{\rm g}\right)$, we obtain $\epsilon$ from Equation~(\ref{eq:h_particle_app}):
\begin{eqnarray}
 && \epsilon \equiv \rho_{\rm d} / \rho_{\rm g} = Z \left(1+\frac{\tau_{\rm stop}}{\alpha}\right)^{1/2}.
  \label{eq:def_ep}
\end{eqnarray}
While $Z$ is a function of $r$ in more realistic situations, we assume $Z$ as a constant in the solar nebula for simplicity.

\subsection{Upper and lower limit of the disk metallicity}
\label{sec:stre-inst}

Since $v_{{\rm d,}r}$ is a decreasing function of $\epsilon$ (see Equation~(\ref{eq:NSH})), a high $\epsilon$ environment is favorable for chondrules to survive.
The high $\epsilon$, however, may trigger the formation of planetesimals via streaming instability (SI).
The SI is a two-fluid instability driven by drifting dust particles in protoplanetary disks \citep{Youdin2005a}.
When $\epsilon$ is high enough, the SI can form dense dust clumps.
These dense dust clumps induce self-gravitational collapse and lead to the formation of planetesimals \citep[e.g.,][]{2012A&A...537A.125J}.
This is one of the most promising processes of the planetesimal formation and occurs quickly under the high $\epsilon$ environment.
Since there is a time gap between the formation of chondrules and parent bodies of meteorites,
the rapid formation of planetesimals is inconsistent with the survival of chondrules in the solar nebula.

Numerical experiments revealed the suitable condition for strong clumping by the SI \citep[e.g,][]{Bai2010b,Bai2010c}.
That condition is formulated as $\epsilon \gtrsim 1$ with relatively large Stokes number, $\tau_{\rm stop} \gtrsim 0.1$ \citep{2014A&A...572A..78D,2016A&A...594A.105D}.
Although the Stokes number of chondrules is much smaller than the conditions shown above, \citet{2015A&A...579A..43C} and \citet{2017A&A...606A..80Y} showed that those small dust particles also can drive the SI in longer timescales than previous SI simulations, $\sim 10^{3}-10^{4}$ yr at $1$ au, which is still significantly shorter than the required lifetime of chondrules.
\edit2{
\citet{2021ApJ...919..107L} conducted similar simulations for more expansive parameter space and found that the critical value of $\epsilon$ for strong clumping weakly depends on the Stokes number of dust particles.
Thus, we regard the critical value of $\epsilon$ as actually an order of unity, even for the small dust particles.
We employ the classical value of $1$ as the critical $\epsilon$ of rapid planetesimal formation via the SI for chondrules in the present study for simplicity.
}

\begin{figure}[ht]
\begin{center}
\includegraphics[width=7cm]{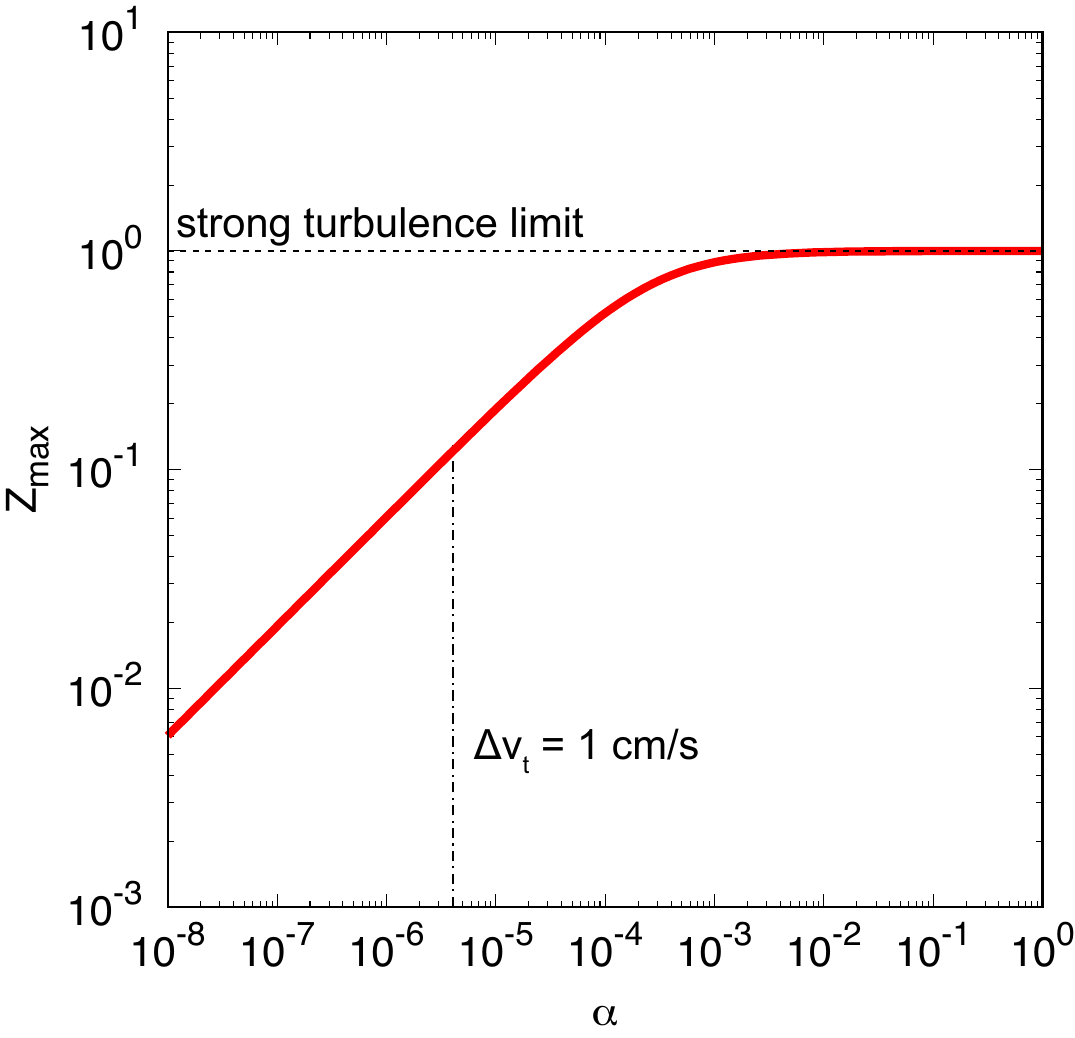}
 \caption{
Maximum disk metallicity $Z_{\rm max}$ as a function of the turbulent viscosity, $\alpha$.
The radii of dust particles are fixed at $1$ mm.
The vertical dot-dashed line shows $\alpha \approx 4 \times 10^{-6}$, which corresponds to $\Delta v_{\rm t} = 1$ cm/s where $\Delta v_{\rm t}$ is the relative velocity between the dust particles driven by the disk turbulence (see Section~\ref{sec:relat-veloc-betw}).
}
 \label{fig:Z_max}
\end{center}
\end{figure}

We define the maximum disk metallicity, $Z_{\rm max}$, which is the disk metallicity corresponding to $\epsilon = 1$ at a given $\alpha$.
Substituting $\epsilon = 1$ into Equation~(\ref{eq:def_ep}), we obtain $Z_{\rm max}$:
\begin{eqnarray}
 Z_{\rm max} = \left(1+ \frac{\tau_{\rm stop}}{\alpha} \right)^{-1/2}.
\label{eq:Z_max}
\end{eqnarray}
Figure~\ref{fig:Z_max} shows $Z_{\rm max}$ as a function of $\alpha$ for $1$ mm dust particles.
When $\alpha \gg \tau_{\rm stop}$, $Z_{\rm max}$ converges to the unity.
This is because the dust scale height agrees with the gas scale height in this regime (Equation~(\ref{eq:h_particle_app})).
When $\alpha \ll \tau_{\rm stop}$, $Z_{\rm max}$ is proportional to $\alpha^{1/2}$.
In this regime, $\epsilon$ is enhanced by the sedimentation of dust particles because of the weak turbulence.
Therefore $\alpha$ limits $Z_{\rm max}$ to prevent the dust particles from the planetesimal formation via the SI.

We estimate the minimum disk metallicity $Z_{\rm min}$ with diagnoses from the chondritic meteorites.
The chondrites consist of chondrules (small particles) and matrix (fine grains).
By virtue of the small size of fine grains, we can assume that they are coupled with gas in a strict sense.
We can ignore the contribution of such fine grains to the surface density of drifting dust particles, $\Sigma_{\rm d}$.
The matrix abundance in chondrites ranges from 20 to 80 vol\% \citep[e.g.,][]{2014mcp..book...65S}
\edit1{
\footnote{
The CI chondrule has the lowest chondrule abundance ($< 5 \%$).
This is because they have experienced aqueous alteration heavily.
As it is hard to recognize the chondrule, we ignore the CI chondrites upon estimating $\chi_{\rm cnd}$.
}}
.
\edit1{
Assuming that the internal density of chondrules equals to that of matrix, we obtain $\Sigma_{\rm d} \gtrsim \chi_{\rm cnd} \Sigma_{\rm d, MMSN}$, where $\Sigma_{\rm d, MMSN}$ is the dust surface density in the MMSN model, and $\chi_{\rm cnd}$ represents the ratio of chondrules to the surface density of the whole solid component.
Now we assume that $0.2 \lesssim \chi_{\rm cnd} \lesssim 0.8$.
}
Here we consider that $\Sigma_{\rm d, MMSN}$ is the minimum value of total surface density of dust components (i.e., the minimum value of a sum of the surface density of chondrules and matrix) to reproduce the current solar system.
Under these assumptions, we restrict the disk metallicity as
\edit1{
\begin{eqnarray}
Z \gtrsim \chi_{\rm cnd} f^{-1} Z_{\rm MMSN},
\label{eq:Z_min_ineq}
\end{eqnarray}
}
where $Z_{\rm MMSN} \equiv \Sigma_{\rm d, MMSN}/{\Sigma_{\rm g, MMSN}}$ is the metallicity in the MMSN model and $f \equiv \Sigma_{\rm g}/\Sigma_{\rm g, MMSN}$ is the factor to examine the disk mass compared with the MMSN.
We employ $f=1$ as a fiducial case.
We note that $\Sigma_{\rm g}$ is significantly larger than the surface density of fine grains.
Therefore we assume that fine grains do not contribute to $\Sigma_{\rm g}$ in Equation~(\ref{eq:Z_min_ineq}).
Henceforth we employ $Z_{\rm min} = 0.8Z_{\rm MMSN}$ for our reference value.

We also introduce $\alpha_{\rm min}$ which is the turbulent strength corresponding to $\epsilon = 1$ at $Z=Z_{\rm min}$.
Substituting $\epsilon = 1$ and $Z=Z_{\rm min}$ into Equation~(\ref{eq:def_ep}), we obtain
\begin{eqnarray}
 \alpha_{\rm min} = \frac{Z_{\rm min}^2}{1-Z_{\rm min}^2} \tau_{\rm stop},
\label{eq:alpha_min}
\end{eqnarray}
which corresponds to the smallest $\alpha$ that can be kept without causing the SI for all $Z > Z_{\rm min}$.

\section{Result}
\label{sec:result}

\subsection{Survival condition for small dust particles}
\label{sec:surv-cond-small}

We derive the survival condition for small dust particles in the solar nebula.
We take the disk metallicity, $Z$, and the turbulent strength, $\alpha$, as input parameters.
Since small dust particles migrate toward the central star at the radial drift velocity (Equation~(\ref{eq:NSH})), their lifetime is estimated as Equation~(\ref{eq:drift_time}).
On the contrary, chondrules demand the minimum lifetime, which is based on the evidence from chondritic meteorites; (1) their wide scattering of formation ages in a single chondrite, (2) \edit1{the discrepancy in the formation age between chondrules and chondritic parent bodies} (see Section \ref{sec:intro}).
We name such required minimum lifetime as a survival time, $t_{\rm surv}$, and take $t_{\rm surv} = 10^6$ yr as a fiducial case.
Hereafter we regard the survival condition of chondrules, $t_{\rm drift} \geq t_{\rm surv}$.

According to Equation~(\ref{eq:drift_time}) with ignoring $\tau_{\rm stop}^2$, the survival condition is given by
\begin{eqnarray}
  \eta \leq \frac{1}{2}\tau_{\rm surv}^{-1}\tau_{\rm stop}^{-1}\left[1+Z\left(1+\frac{\tau_{\rm stop}}{\alpha}\right)^{1/2}\right]^2,
 \label{eq:survival_condition}
\end{eqnarray}
where all timescales $t_{\ast}$ are normalized as $\tau_{\ast} \equiv t_{\ast}\Omega$.
The other expression of the same condition is
\begin{eqnarray}
 Z \geq \left[\left(2\tau_{\rm surv} \tau_{\rm stop} \eta \right)^{1/2} -1 \right] \left(\frac{\tau_{\rm stop}}{\alpha} + 1 \right)^{-1/2}.
  \label{eq:survival_condition_Z}
\end{eqnarray}
Equations~(\ref{eq:survival_condition}) and (\ref{eq:survival_condition_Z}) represent that small $\eta$ and/or large $Z$ are suitable conditions for small dust particles to survive.
This is because the radial drift velocity of dust particles becomes small under such situations.

We now consider the critical case, corresponding to $t_{\rm surv} = t_{\rm drift}$.
The critical dimensionless pressure gradient force, $\eta_{\rm c}$, can be given as
\begin{eqnarray}
 \eta_{\rm c} = \frac{1}{2}\tau_{\rm surv}^{-1}\tau_{\rm stop}^{-1}\left[1+Z\left(1+\frac{\tau_{\rm stop}}{\alpha}\right)^{1/2}\right]^2.
  \label{eq:eta_crit}
\end{eqnarray}
Since Equation~(\ref{eq:eta_crit}) is a quadratic equation in $Z$, we have the minimum value of $\eta_{\rm c}$: $\eta_{\rm c, min} \equiv \tau_{\rm stop}^{-1} \tau_{\rm surv}^{-1} /2$.
As we take $\epsilon = 1$ as the threshold condition for planetesimal formation (see Section \ref{sec:stre-inst}), we have the maximum disk metallicity $Z_{\rm max}$ (Equation~(\ref{eq:Z_max})).
Substituting $Z=Z_{\rm max}$ in Equation~(\ref{eq:eta_crit}), we derive the maximum value of $\eta_{\rm c}$ as $\eta_{\rm c, max} \equiv 2\tau_{\rm stop}^{-1} \tau_{\rm surv}^{-1}$.
Note that these minimum and maximum values of $\eta_{\rm c}$ are independent of $\alpha$ and $r$, because
\begin{eqnarray}
 \tau_{\rm stop}^{-1}\tau_{\rm surv}^{-1} \approx 5.8 \times 10^{-4} \left(\frac{a}{1 \, {\rm mm}}\right)^{-1}\left(\frac{t_{\rm surv}}{10^{6} \, {\rm yr}}\right)^{-1}.
\label{eq:tautau}
\end{eqnarray}
In other words, they only depend on requirements from the material properties of the chondrules, not on the properties of the solar nebula (Equation~(\ref{eq:tautau})).
Therefore, these values can be treated as constants when considering the survivability of chondrules by varying only the disk parameters.
Using $Z_{\rm max}$ and $\eta_{\rm c, min}$, we have a straightforward formula of $\eta_{\rm c}$ as
\begin{eqnarray}
 \eta_{\rm c} = \eta_{\rm c, min}\left(1+\frac{Z}{Z_{\rm max}}  \right)^2.
\label{eq:eta_crit_rewrite}
\end{eqnarray}

Here we consider two extreme cases which \edit1{correspond} to the upper and lower limit of $\alpha$.
As we take $Z_{\rm max} = 1$ in the case of strong turbulence ($\alpha \gtrsim 10^{-2}$, see Figure {\ref{fig:Z_max}}), $\eta_{\rm c}$ is given by,
\begin{eqnarray}
 \eta_{{\rm c}, \alpha_{\infty}} = \eta_{\rm c, min}\left(1+Z\right)^2.
\label{eq:eta_crit_alpha_inf}
\end{eqnarray}
Conversely, more dust particles sediment into the disk midplane under the weak turbulence.
Under such conditions, $Z_{\rm max}$ agrees with $Z_{\rm min}$, so that we finally have the condition of $\eta_{\rm c}$ under the weak turbulence as
\begin{eqnarray}
 \eta_{{\rm c}, \alpha_{\rm min}} = \eta_{\rm c, min}\left(1+\frac{Z}{Z_{\rm min}}\right)^2.
\label{eq:eta_crit_alpha_min}
\end{eqnarray}

\edit1{
In summary, we have the region bounded by $Z_{\rm min}$, $\eta_{\rm c, max}$, and $\eta_{{\rm c}, \alpha_{\infty}}$ as
}
\begin{eqnarray}
\left\{ \begin{array}{l}
Z_{\rm min} \leq Z \leq 1, \\
\eta_{\rm c, min}(1+Z)^2 \leq \eta_{\rm c} \leq \eta_{\rm c, max}. \\
\end{array}
\right.
\label{eq:condition_Z-eta_plane}
\end{eqnarray}
\edit1{
This is a set of all $\eta$ that are the closest to the MMSN value among the disks in which the chondrules can survive for given $Z$ and $\alpha$.
We note that Equation~(\ref{eq:condition_Z-eta_plane}) depends on the values of chondrules' size and survival time from material evidence of chondritic meteorites.
}
In the following subsections, we represent the survival condition for chondrules parameterizing the size of chondrules and the survival time.

\subsection{Results of the fiducial case}
\label{sec:results-fiduc-model}

\begin{figure}[ht]
\begin{center}
\includegraphics[width=7cm]{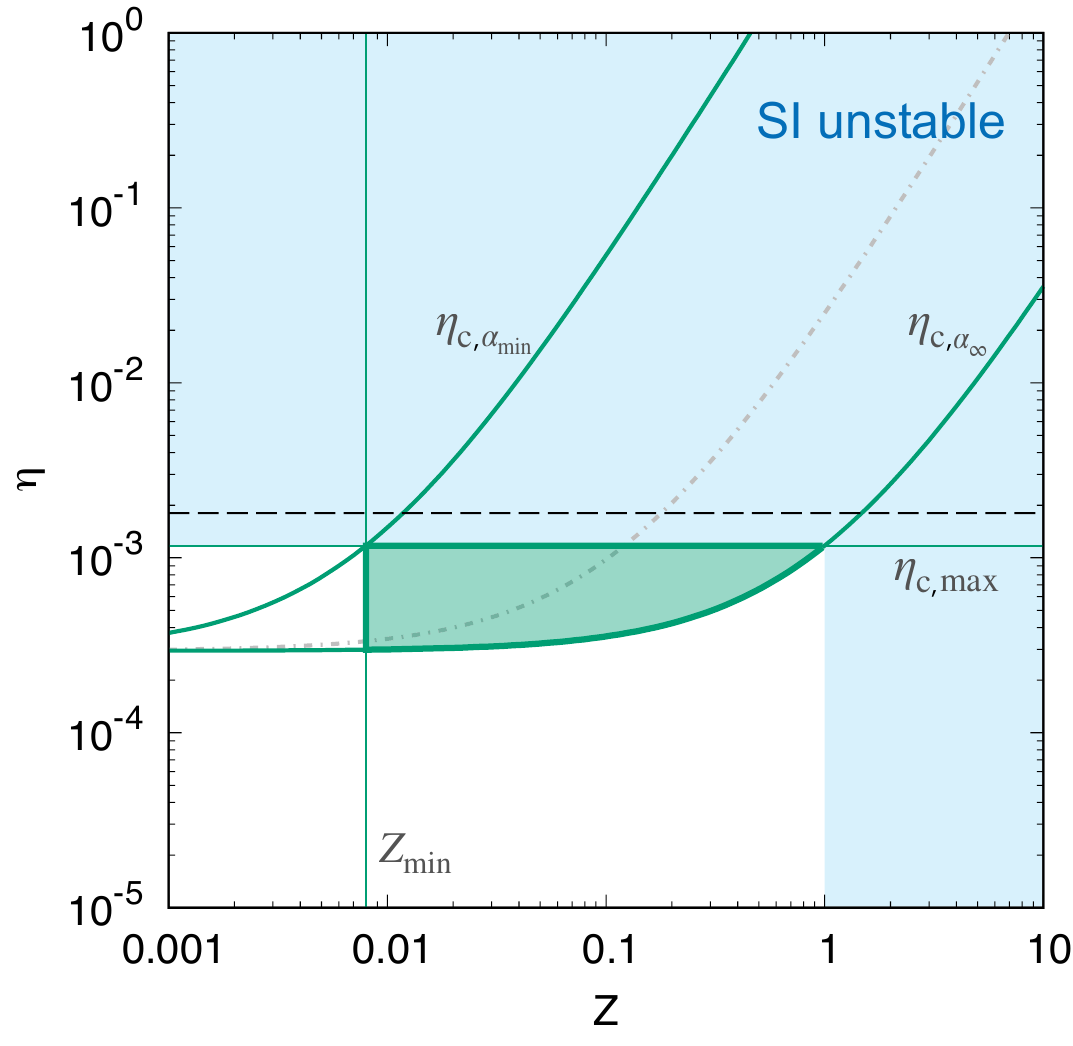}
 \caption{
$\eta_{\rm c}$ as a function of $Z$ for our fiducial case.
\edit1{
The green shaded region surrounded by thick green lines corresponds to the region of Equation~(\ref{eq:condition_Z-eta_plane}) for our fiducial case.
}
The horizontal green solid line depicts $\eta_{\rm c, max}$ and the horizontal gray dashed line is $\eta_{\rm MMSN}$.
Two green solid curves represent two extreme cases of the lower (left) and upper (right side) limit of turbulent strength.
The blue-colored area represents the SI unstable region.
The gray dot-dashed curve depicts $\eta_{\rm c}$ with $\alpha=4\times 10^{-6}$, which corresponds to $\Delta v_{\rm t} = 1.0 \ {\rm cm}/{\rm s}$ (see Section~\ref{sec:relat-veloc-betw}).
}
 \label{fig:eta_crit}
\end{center}
\end{figure}

Figure~\ref{fig:eta_crit} depicts $\eta_{\rm c}$ as a function of $Z$ in our fiducial case of $a=1$ mm and $t_{\rm surv} = 10^{6}$ yr.
Note that we fix $r=1$ au, $Z_{\rm MMSN}= 0.01$, and $f=1$ in Section~\ref{sec:result}.
\edit1{
The green-colored area is the region represented by Equation~(\ref{eq:condition_Z-eta_plane}) (hereafter, we call this half semi-circle-like shaped region the ``shaded region'').
In the blue-colored area, the SI becomes unstable, and planetesimals form.
}
The horizontal green solid line depicts $\eta_{\rm c, max}$ which is given by
\begin{eqnarray}
 \eta_{\rm c, max} \approx 1.17 \times 10^{-3} \left(\frac{a}{1 \, {\rm mm}}\right)^{-1} \left(\frac{t_{\rm surv}}{10^{6} \, {\rm yr}}\right)^{-1}.
\label{eq:eta_max_normal}
\end{eqnarray}
We find that $\eta_{\rm c, max}$ is slightly smaller than $\eta_{\rm MMSN}$ (the horizontal gray dashed line in Figure~\ref{fig:eta_crit}).
Therefore, chondrules can survive for $10^{6}$ yr in the relatively flatter solar nebula than the nominal MMSN disk.
We note that $\eta_{\rm c, min}$ is one order of magnitude smaller than $\eta_{\rm MMSN}$,
\begin{eqnarray}
 \eta_{\rm c, min} \approx 2.94\times 10^{-4} \left(\frac{a}{1 \, {\rm mm}} \right)^{-1}\left(\frac{t_{\rm surv}}{10^{6} \, {\rm yr}} \right)^{-1}.
\label{eq:eta_min_norm}
\end{eqnarray}

We show two cases with different turbulent \edit1{strengths} as green solid curves in Figure~\ref{fig:eta_crit}:
the left curve depicts $\eta_{{\rm c}, \alpha_{\rm min}}$ (Equation~(\ref{eq:eta_crit_alpha_min})) which corresponds to the weak turbulent case, and the right curve is  $\eta_{{\rm c}, \alpha_{\infty}}$ (Equation~(\ref{eq:eta_crit_alpha_inf})) which corresponds to the strong turbulent case.
From Equation~(\ref{eq:alpha_min}), the minimum value of $\alpha$ is
\begin{eqnarray}
 \alpha_{\rm min} \approx 6.0 \times 10^{-5} \tau_{\rm stop} \approx 2.0 \times 10^{-8}.
\end{eqnarray}
When we take the nominal MMSN values ($Z_{\rm MMSN} \approx 0.01$ and $\eta_{\rm MMSN} \approx 10^{-3}$), the disk turbulence should be $\alpha \sim 10^{-8}$, which is significantly weaker than the typical value of the protoplanetary disk models, $\alpha \sim 10^{-2} $\citep{1998ApJ...495..385H}.
On the other hand, $\eta_{\rm c}$ for the case with $\alpha \sim 10^{-2}$ is almost identical to $\eta_{{\rm c},\alpha_{\infty}}$.
If the disk metallicity is similar to the MMSN value ($Z_{\rm MMSN} \approx 0.01$), then $\eta$ needs to be almost less than $\eta_{\rm c, min}$.
In this case, the solar nebula should have an inverted surface density distribution, which is quite characteristic (see Section~\ref{sec:plaus-disk-struct}).

\edit1{
We note that the shaded region corresponding to Equation~(\ref{eq:condition_Z-eta_plane}) is independent of the initial orbital radius as we expressed in Equations~(\ref{eq:tautau}) and (\ref{eq:condition_Z-eta_plane}).
}
This is because we assume that $t_{\rm drift}$ is free from the orbital radius (Equation~(\ref{eq:fiducial_drift_time})).
We discuss \edit1{the} validity of this assumption in Section~\ref{sec:radial-location-cv}.

\begin{figure}[ht]
\begin{center}
\includegraphics[width=8.6cm]{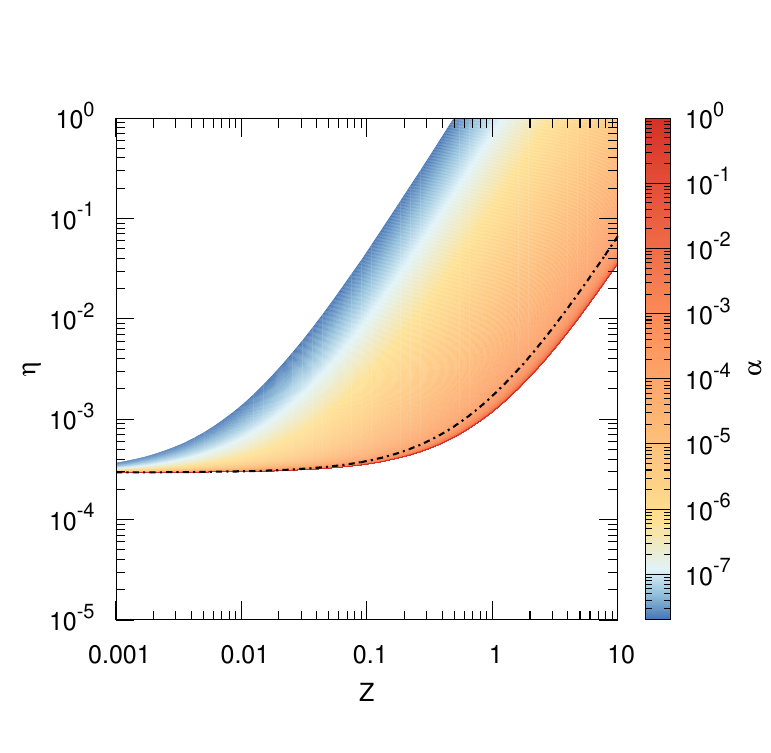}
 \caption{
\edit2{
Dependence of $\eta_{\rm c}$ on $\alpha$ for our fiducial case.
The color illustrates $\alpha$, which ranges from $2.0 \times 10^{-8}$ to $10^{0}$.
The dot-dashed line depicts the case of $\alpha = \tau_{\rm stop} (\approx 10^{-4})$.
}}
 \label{fig:contour}
\end{center}
\end{figure}

\edit2{
We also check the dependence of $\eta_{\rm c}$ on $\alpha$ for our fiducial case.
Considering a wide range of $\alpha$ (from $2 \times 10^{-8}$ to $1$, the same as in Figure~\ref{fig:Z_max}), we plot the corresponding $\eta_{\rm c}$ in Figure~\ref{fig:contour}.
When $\alpha \gtrsim \tau_{\rm stop} (\approx 10^{-4})$, $\eta_{\rm c}$ is almost stable and the range of $\eta_{\rm c}$ and $Z$ for surviving chondrules is narrow.  
This is because the dust scale height easily approaches the gas scale height due to the turbulent stirring (see Section~\ref{sec:dust-gas-mass}).  
On the opposite side, when $\alpha \lesssim \tau_{\rm stop}$, the range of $\eta_{\rm c}$ and $Z$ becomes wide. 
Furthermore, the effect of the gas accretion on the chondrule lifetime becomes important only in the region below the dot-dashed line in Figure~\ref{fig:contour} ($\alpha = \tau_{\rm stop}$).
As this region is very limited, this also confirms that the effect of the gas accretion is less effective as we mentioned in Section~\ref{sec:radi-drift-timesc} and Appendix~\ref{appendixA}. 
Overall, our favorable condition in the disk for surviving chondrules are $\alpha \ll \tau_{\rm stop}$, where the gas accretion becomes inefficient to the dust radial velocity.
}

\edit1{
\subsection{Effects of particles size}
}
\label{sec:depend-radi-dust}

Chondrules and CAIs have a considerable size variation of the two orders of magnitude (see Section~\ref{sec:intro}).
Since the radial drift velocity is an increasing function of the Stokes number at $\tau_{\rm stop} < 1$ (Equation~(\ref{eq:NSH})), it is important to examine the dependence on the radius of dust particles, $a$.

Smaller particles naturally have a longer lifetime because of their small radial drift velocity.
When $\eta_{\rm c, min}$ is larger than $\eta_{\rm MMSN}$, the dust particles can survive for $t_{\rm surv}$ even in MMSN-like disks, which have the same radial profile of temperature and density as that for MMSN.
The ratio of $\eta_{\rm c, min}$ to $\eta_{\rm MMSN}$ from Equations~(\ref{eq:eta_mmsn}) and (\ref{eq:eta_min_norm}) indicates such the condition,
\begin{eqnarray}
 \frac{\eta_{\rm c, min}}{\eta_{\rm MMSN}} \approx 0.16 \left( \frac{a}{1 \, {\rm mm}}\right)^{-1} \left( \frac{t_{\rm surv}}{10^{6} \, {\rm yr}}\right)^{-1}.
\label{eq:eta-ratio}
\end{eqnarray}
When dust particles are smaller than $0.16$ mm in radius, they can survive $10^{6}$ yr in MMSN-like disks.

\begin{figure}[ht]
\begin{center}
\includegraphics[width=7cm]{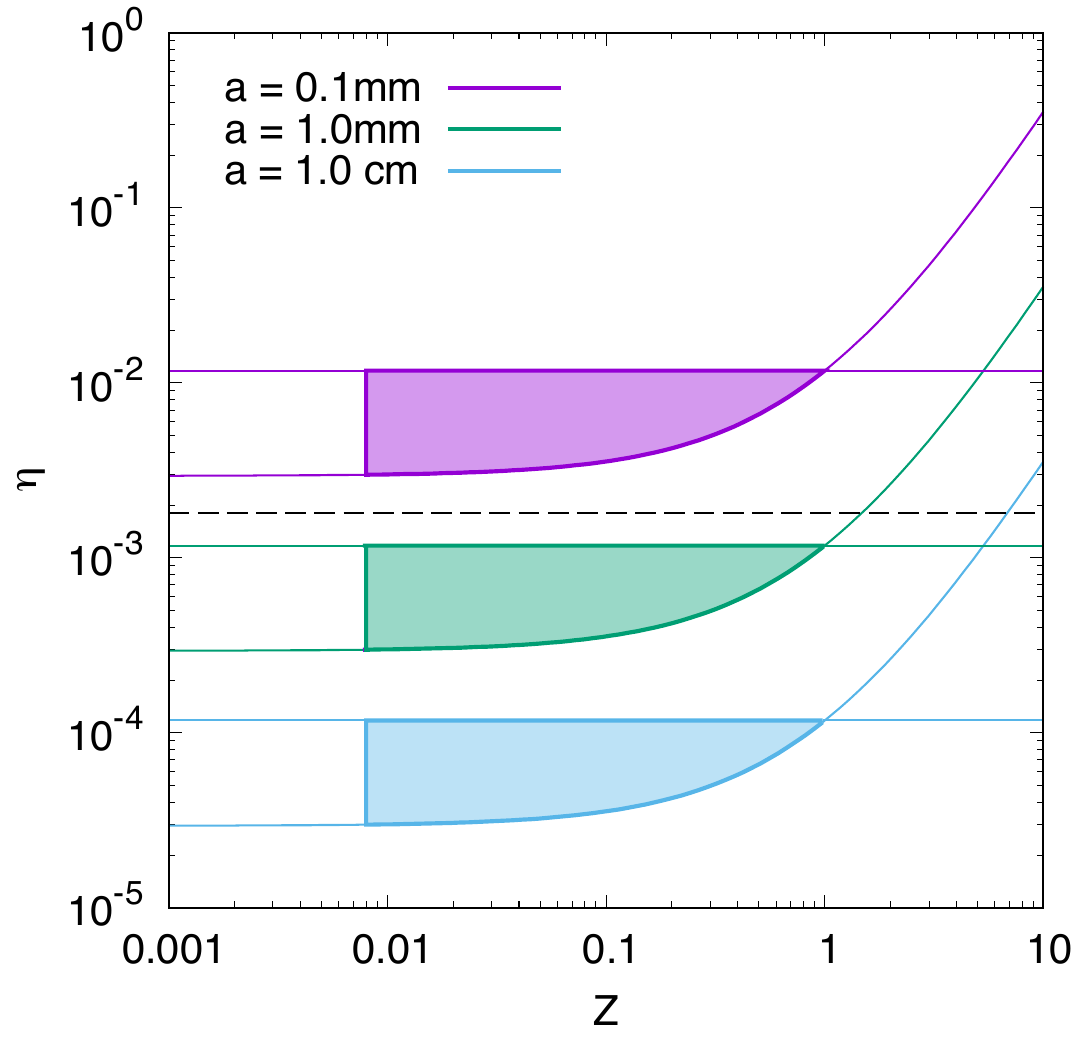}
 \caption{
\edit1{
Shaded regions corresponding to Equation~(\ref{eq:condition_Z-eta_plane}) for different particle sizes: $a=0.1$ mm (purple), $a=1.0$ mm (green; fiducial case), and $a=1.0$ cm (light blue), respectively.
}
Each colored line represents $\eta_{\rm c, max}$ (horizontal lines) and $\eta_{{\rm c,} \alpha_{\infty}}$ (curved lines) for each case.
}
 \label{fig:radius}
\end{center}
\end{figure}

\edit1{
The shaded region corresponding to Equation~(\ref{eq:condition_Z-eta_plane}) also has a considerable variation in the $\eta$-$Z$ plane because of the large variety of particle sizes.
Figure~\ref{fig:radius} illustrates the shaded regions for different particles radii.
The purple-colored area in Figure~\ref{fig:radius} depicts the shaded region for $0.1$ mm sized particles when $t_{\rm surv} = 10^{6}$.
}
In this case, as expected by Equation~(\ref{eq:eta-ratio}), $\eta_{\rm c}$ always exceeds $\eta_{\rm MMSN}$.
Thus, relatively small chondrules of $0.1$ mm as in CO and CH chondrites \citep[e.g.,][]{2014mcp..book...65S}, can survive easier than $1$ mm sized nominal chondrules even in the MMSN-like disks.

Although some CAIs have an order of magnitude larger radius than nominal chondrules, these CAIs also should survive for $10^{6}$ yr in the solar nebula.
\edit1{
The light blue area in Figure~\ref{fig:radius} is the shaded region for $1$ cm sized particles.
}
In this case, $\eta \lesssim 10^{-4}$ is required and this $\eta$ value is significantly smaller than $\eta_{\rm MMSN}$ (Equation~(\ref{eq:eta_mmsn})).
The survivability of CAIs is more severe than the chondrules.

In order for dust particles of all sizes to survive equally, the most stringent conditions \edit1{must} be achieved where even the shortest lived particles can survive.
We can naively consider that the solar nebula has an extremely small $\eta$ where CAIs (largest particles) can survive for $\sim 10^{6}$ yr against the radial drift.
However, CAIs are relatively minor components than chondrules.
Perhaps the survival mechanisms of chondrules and CAIs should be considered separately, i.e., there is some additional \edit2{mechanisms} to elongate CAIs lifetime \edit2{
\citep[e.g., outward radial diffusion processes;][]{2003Icar..166..385C,2018ApJS..238...11D,2022SciA....8M3045L}
}
\footnote{
Note that dust trapping in the planetary gap edge created by Jupiter plays an essential role in the CAI survival scenario of \citet{2018ApJS..238...11D}.
\edit2{
The outward migration of small particles can also increase the dust density locally \citep[see][]{2022SciA....8M3045L}.
We discuss the effects of such dust accumulation processes in Section~\ref{sec:dust-trap}.
}
We also discuss a possible scenario that satisfies CAI storage due to the dust trapping and chondrule survival with a delayed radial drift in Section~\ref{sec:variation-gas-disk}.
}.

\edit1{
\subsection{Effects of survival time}
}
\label{sec:depend-surv-time}

\edit1{
Equation~(\ref{eq:condition_Z-eta_plane}) has a same dependency on the particle size, $a$, and $t_{\rm surv}$ (see Equation~(\ref{eq:tautau})).
}
The formation of chondrules starts just after the formation of CAIs and continues to $5$ Myr after that \citep[e.g.,][]{cbk12,bbk15}.
The variation of chondrule ages in a single chondrite \citep{cbk12,2009Sci...325..985V} and the time gaps between the formation of chondrules and the accretion of chondritic planetesimals are on the order of $10^{6}$ yr \citep[e.g.,][]{dac14,2014prpl.conf..571G}.
Therefore, we adopt $t_{\rm surv} = 10^{6}$ yr as our fiducial case.
Although the variation of $t_{\rm surv}$ is relatively minor than that of the particle size, $a$,
$t_{\rm surv}$ has an uncertainty of a factor of a few. 

\begin{figure}[ht]
\begin{center}
\includegraphics[width=7cm]{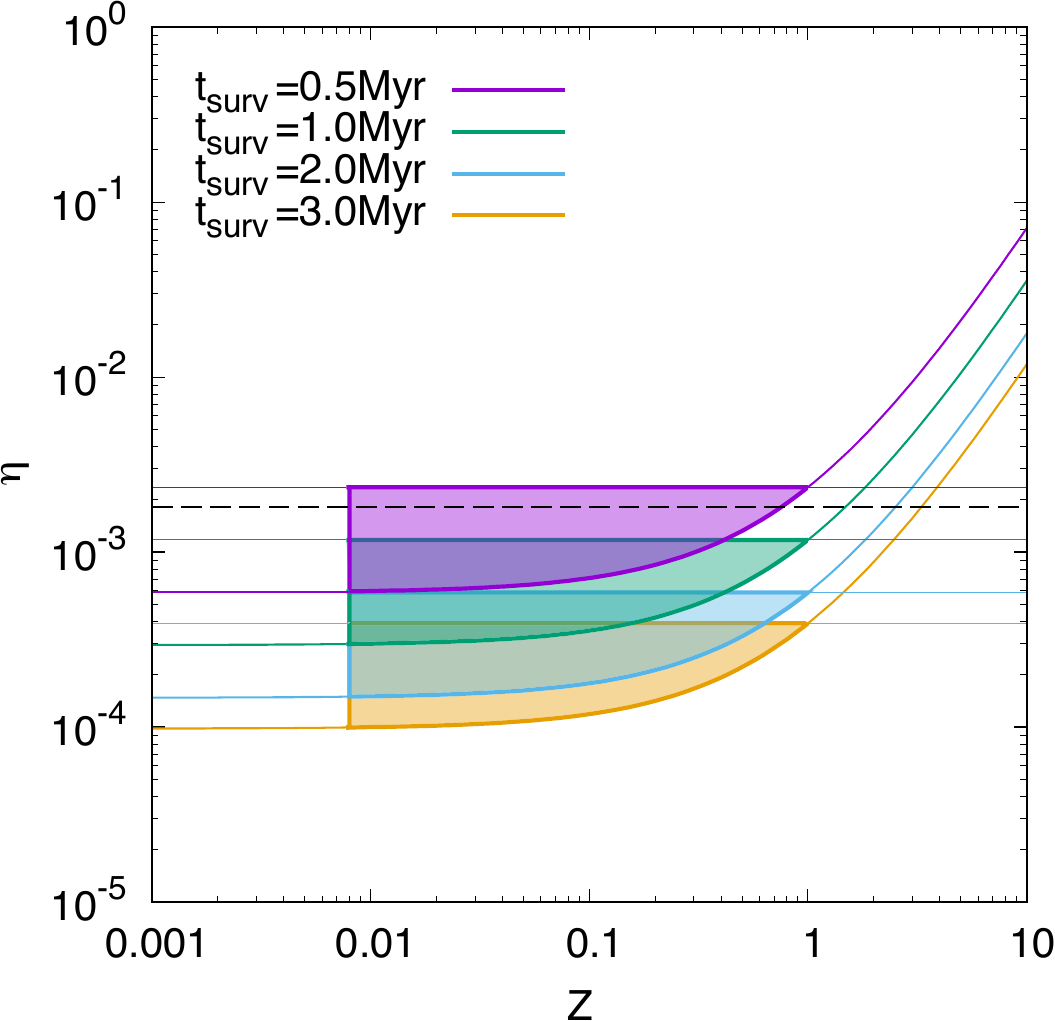}
 \caption{
\edit1{
Shaded regions corresponding to Equation~(\ref{eq:condition_Z-eta_plane}) for different survival times: $t_{\rm surv} = 5\times 10^{5}$ yr (purple), $t_{\rm surv} = 1 \times 10^{6}$ yr (green; fiducial case), $t_{\rm surv} = 2\times 10^{6}$ yr (light blue), and $t_{\rm surv} = 3\times 10^{6}$ yr (orange), respectively.
}
Each colored line represents $\eta_{\rm c, max}$ (horizontal lines) and $\eta_{{\rm c,} \alpha_{\infty}}$ (curved lines) for each case.
}
 \label{fig:t_disk}
\end{center}
\end{figure}

\edit1{
Here we check the effects of $t_{\rm surv}$.
Figure~\ref{fig:t_disk} shows the shaded regions, which correspond to Equation~(\ref{eq:condition_Z-eta_plane}), for different survival times.
The shaded region for the case with $t_{\rm surv} = 5\times 10^{5}$ yr (purple) overlap $\eta_{\rm MMSN}$, thus the MMSN model can work for this case.
For the case with $t_{\rm surv} = 3\times 10^{6}$ yr (orange), $\eta_{\rm c, max} \approx 4 \times 10^{-4}$ is required; this is an order of magnitude smaller $\eta$ than the MMSN model.
}

\section{Discussion}
\label{sec:discussion}

\subsection{Conditions for not growing in size}
\label{sec:chondr-should-keep}
We discuss the validity of our assumption: small particles escape from further growth and/or destruction during their survival time in the solar nebula.
Since there should be a time gap between the formation time of chondrules and that of planetesimals, chondrules keep their original size during $t_{\rm surv}$.
We consider two effects which potentially induce size evolution of chondrules, (1) pairwise collision of particles, and (2) accumulation of particles in radial/azimuthal direction in the solar nebula.

\subsubsection{Pairwise collision of particles}
\label{sec:relat-veloc-betw}

We discuss the size evolution of small particles due to the mutual collision.
The efficiency of size evolution is controlled by the relative velocity of colliding particles.
The relative velocity of small particles in protoplanetary disks, $\Delta v_{\rm pp}$, is expressed as \citep[e.g.,][]{Okuzumi2012a}
\begin{eqnarray}
\Delta v_{\rm pp} = \sqrt{ \left(\Delta v_{\rm B}\right)^2 + \left(\Delta v_{\rm mean}\right)^2 + \left(\Delta v_{\rm t}\right)^2}.
\end{eqnarray}
where $\Delta v_{\rm B}$ is the relative velocity driven by the Brownian motion, $\Delta v_{\rm mean} = \sqrt{\left(\Delta v_{\rm r}\right)^2 + \left(\Delta v_{\rm \phi}\right)^2 + \left(\Delta v_{\rm z}\right)^2}$ is the relative velocity due to the mean flow in each direction, and $\Delta v_{\rm t}$ is the \edit2{relative} velocity \edit2{driven} by the gas turbulence.
The Brownian motion only has a small contribution to $\Delta v_{\rm pp}$ for chondrule-size particles.
Contributions to $\Delta v_{\rm pp}$ from the mean flow and turbulence are roughly comparable for chondrule-size particles \citep[e.g.,][]{Sato2016a}.
Thus, we assume that $\Delta v_{\rm pp} \sim \Delta v_{\rm t}$ for simplicity\footnote{This paper claims that the disk with small $\eta$ may be favorable for the chondrule surviving. The contribution to $\Delta v_{\rm pp}$ from the mean flow in $r$ and $\phi$ direction naturally becomes negligible under such situations.}.

\edit1{
We can estimate $\Delta v_{\rm t}$ for chondrules as
\begin{eqnarray}
   \Delta v_{\rm t} \approx \left\{\begin{array}{ll}
\delta v_g {\rm Re}_t^{1/4} C_{1} t_{\rm cnd}\Omega, & \hspace{0.5cm} t_{\rm cnd} \ll t_\eta, \\
\delta v_g\sqrt{C_{2} t_{\rm cnd}\Omega},  & \hspace{0.5cm} t_\eta \ll t_{\rm cnd} \ll \Omega^{-1}, \end{array} \right. &\label{eq:dvt}
\end{eqnarray}
where $\delta v_{g} = \sqrt{\alpha}c_{\rm s}$ is the gas velocity driven by the turbulence, ${\rm Re}_{t} = \alpha c_{\rm s} h_{\rm g}/\nu_{\rm m}$ is the turbulent Reynolds number with a kinematic viscosity of the gas $\nu_{\rm m} = \lambda_{\rm mfp} v_{\rm th}/2$, $t_{\eta} = {\rm Re}_{t}^{-1/2} \Omega^{-1}$ is the turnover time of the smallest eddies, $C_{1}$ ($C_{2}$) is a factor from the size ratio of the colliding chondrules for the first (second) regime, and $t_{\rm cnd}$ is the stopping time of chondrules \citep[see Equations (27) and (28) of][]{Ormel2007a}.
We choose $C_{1} = 1/2$ and $C_{2} = 3$ to include the effect of size distribution of chondrules \citep{Sato2016a}.
}
We note that  $\Delta v_{\rm t}$ is an increasing function of $\tau_{\rm stop}$ when $\tau_{\rm stop} \leq 1$.

Dust particles usually have the maximum size when the relative velocity is equal to their critical velocity ($v_{\rm c}$), which is the minimum velocity when colliding particles no longer grow into larger ones.
However, the critical velocity of chondrules is uncertain.
We sometimes observe chondrules and CAIs with surrounding fine grained rims (e.g., silicate particles), which is apparently different from the matrix \citep[e.g.,][]{2014mcp..book...65S}.
This indicates that they acquired their rims during in the solar nebula; after their formation but before they embedded in their parent body \citep[e.g.,][]{Cuzzi:2004,Hanna:2018,Simon:2018}.
Therefore we assume that the critical velocity of chondrules (or aggregates which include them) is similar to the critical velocity of same size aggregates which consist of micron-sized silicate particles (i.e., monomers).
Although the fine grains attached at the surface of chondrules increase the size of chondrules, we assume that the fine-grained rims do not change the Stokes number of core particles for simplicity.

The critical velocity of silicate aggregates is studied experimentally and numerically.
Laboratory experiments show that $v_{\rm c} \sim 10 \ {\rm cm/s}-1 \ {\rm m/s}$ for fragmentation of silicate aggregates \citep[e.g.,][]{2010A&A...513A..56G}.
Alternatively, the numerical simulations suggest that $v_{\rm c} \sim 1-10 \ {\rm m/s}$ \citep{2011ApJ...737...36W,Wada2013a}, which is one order of magnitude higher than the experimental one.
The critical velocity of $\sim 1$ cm/s is the other threshold when the bouncing of colliding aggregates occurs \citep{2010A&A...513A..57Z}.
We also note that recent work reported $v_{\rm c} \sim 60-80 \ {\rm m}/{\rm s}$ \citep{Kimura15,Steinpilz19}, which is the one order of magnitude higher critical velocity of silicate grains for fragmentation.
\citet{2016ApJ...832L..19A} also pointed out that the uncertainty of silicate monomer size observed in meteorites may increase the critical velocity of the dust aggregates (i.e., the conditions for chondrule surviving become more severe).
Here we optimistically employ the lowest critical velocity, $v_{\rm c} \approx 1 \ {\rm cm/s}$, which corresponds to the widest parameter space allowing chondrules to survive.

When the relative velocity for the chondrule-size particle is lower than $v_{\rm c}$, the particle will grow beyond the chondrule size.
Therefore, we roughly determine the minimum value of $\alpha$ in terms of inhibition of the growth of chondrules by equating $v_{\rm c}$ to $\Delta v_{\rm t}$ (Equation (\ref{eq:dvt})):
\begin{eqnarray}
 \alpha \approx \left\{\begin{array}{ll}
4 \times 10^{-6} \left(\frac{v_{\rm c}}{1 \, {\rm cm}/{\rm s}}\right)^{4/3}, & \hspace{0.5cm} v_{\rm c} \ll 4.5 \, {\rm cm}/{\rm s}, \\
1.3\times 10^{-7} \left(\frac{v_{\rm c}}{1 \, {\rm cm}/{\rm s}}\right)^{2}, & \hspace{0.5cm} v_{\rm c} \gg 15 \, {\rm cm}/{\rm s}, \\
\end{array} \right.&
\label{eq:alpha-dvt}
\end{eqnarray}
which is $\alpha$ at the critical velocity for $1$ mm particles
\edit1{
\footnote{The two regimes in Equation~(\ref{eq:alpha-dvt}) correspond to the two regimes in Equation~(\ref{eq:dvt}).
When the regimes switch by $t_{\eta}$ in Equation~(\ref{eq:dvt}), the relative velocities are not smoothly connected.
Therefore, the regimes that switch by $v_{\rm c}$ is not continuous in Equation~(\ref{eq:alpha-dvt}).}
}
.
\edit1{
When
}
we set $v_{\rm c} \approx 1 \ {\rm cm/s}$, the minimum value of $\alpha$ is $4\times 10^{-6}$.
The vertical dot-dashed line in Figure~\ref{fig:Z_max} represents $\alpha = 4\times 10^{-6}$ and the gray dotted line in Figure~\ref{fig:eta_crit} shows $\eta_{\rm c}$ with $\alpha = 4\times 10^{-6}$.
\edit1{
As shown in Figure~\ref{fig:eta_crit}, the region where chondrules can survive on the $\eta$-$Z$ plane is narrowed by considering the condition for the collisional velocity.
}
For instance, $Z \gtrsim 0.1$ is needed when we employ $\eta \approx 10^{-3}$.

\subsubsection{Accumulation in radial/azimuthal direction}
\label{sec:dust-trap}

Here we discuss the effects of the particle concentration in the radial/azimuthal direction, which can elongate the lifetime of chondrules.
Additional growth of chondrules and/or planetesimal formation can simultaneously occur because the dust-to-gas mass ratio increases.

The dust particles of constant size start to drift inward in the radial direction of the disk at the same time. 
The dust particles naturally start to concentrate in the inner region of the disk because both $\tau_{\rm stop}$ and $v_{{\rm d}, r}$ decrease there \citep{2002ApJ...580..494Y}.
Besides, when the disk has a substructure, the drifting dust particles are trapped there and enhance the concentration.
These are disk substructures that can concentrate dust particles in the radial direction; pressure bumps \citep{1972fpp..conf..211W,Taki2016a}, the zonal flow formed by the magneto-rotational instability \citep{2009ApJ...697.1269J,2013ApJ...763..117D}, and planetary gaps \citep{2009A&A...493.1125L}.
The large scale vortices which are formed by the Rossby-wave instability \citep{1999ApJ...513..805L} also concentrate dust particles in the radial and azimuthal direction \citep{2009A&A...493.1125L,2014ApJ...795L..39F}.
Regardless of which of the above mechanisms is used to concentrate the dust particles efficiently, the dust-to-gas mass ratio easily reaches unity.
Under such circumstances, planetesimals are expected to form by various pathways (i.e., direct sticking, gravitational instability, and SI).

Since the radial/azimuthal accumulation of dust grains easily induces the subsequent growth of dust particles, these processes of dust concentration are unsuitable for chondrules to survive
\edit1{
as free-floating particles
}
.
Contrary, the formation of local substructures in the late stage of the solar nebula may support the formation of planetesimal from chondrules (see Section~\ref{sec:plan-form-as}).
For instance, dust ring formation associated with the inside-out gas dispersal in protoplanetary disks is one possible candidate for such a mechanism \citep{Suzuki2016a,Takahashi2018a,Taki_2021}.

\subsection{Where is the chondrule formation region?}
\label{sec:radial-location-cv}

The formation location of chondrites and chondrules is unclear.
Though we assume that the chondrules are initially located at a few au from the Sun where the rocky planets are formed there (Section~\ref{sec:result}), oxygen isotopic studies of chondrules may suggest their formation region.
Chondrules in carbonaceous chondrites have different oxygen isotopic compositions from that in ordinary chondrites \citep[e.g.,][]{Tenner:2015,Hertwig:2019}.
This may indicate that chondrule forming locations between carbonaceous and ordinary chondrites also differ.
The radial transport of chondrules is one possible scenario to explain the oxygen isotopic compositions \citep{Yamanobe:2018,Williams:2020}, as in the same context of chondrule-like objects found in the Wild2 comet \citep{Nakamura:2008}.
Moreover, other isotopic studies indicate clear differences between carbonaceous and non-carbonaceous chondrites \citep{Warren:2011,2017PNAS..114.6712K,Kleine:2020}.
Their compositional fields would be separated by a wide gap in the solar nebula.
If these isotopic differences indicate the different \edit1{locations} in the solar nebula, chondrule formation could have occurred at various \edit1{locations} in the solar nebula.

We here discuss that the chondrule-forming region locates in outside of the H$_2$O snow line, where $ r \approx 2.7$ au in the MMSN model.
One problem conceivably caused by H$_2$O ice is the growth of chondrules.
Since the fragmentation velocity of icy grains is about one order of magnitude higher than that of rocky grains, the dust growth model predicts that an equilibrium value of $\tau_{\rm stop}$ increases outside the H$_2$O snow line \citep[e.g.,][]{Birnstiel2010a}.
Such large dust particles easily drive the SI and quickly form the planetesimals \citep{2014A&A...572A..78D}.
In addition, without considering the SI, chondrules covered with an icy fine-grain rim in the solar nebula, should grow into larger bodies due to the direct sticking \citep[e.g.,][]{Okuzumi2012a}.
The timescales of these phenomena are significantly shorter than $10^{6}$ yr, the nominal value of $t_{\rm surv}$.

We also point out that the radial drift timescale is primarily independent of the initial locations of chondrules as long as the size of chondrules is fixed.
As we ignore a dependency of $\eta$ on the orbital radius assuming $r \sim 1$ au (Equation~(\ref{eq:eta_mmsn}) in Section~\ref{sec:radi-drift-timesc}), $t_{\rm drift}$ does not depend on $r$ in a strict sense (Equation~(\ref{eq:fiducial_drift_time})).
To consider a wide orbital range of the disk, we here derive $t_{\rm drift}$ including the dependency of $\eta$ on $r$.
In the MMSN model, $\eta$ is expressed as $\eta = 1.8 \times 10^{-3} \left(r/1 \ {\rm au}\right)^{1/2}$.
Thus, we obtain $|v_{r}| \propto r^{3/2}$ and $t_{\rm drift} \equiv r/|v_{r}| \propto r^{-1/2}$.
The fact that $t_{\rm drift}$ is proportional to the negative power of $r$ means that the total migration timescale of dust particles is dominated by the crossing time in the inner part of the disk.

The crossing time of dust particles only has a weak dependency on their initial location.
We calculate the crossing time of chondrules as
\begin{eqnarray}
&&t_{\rm cross} \equiv \int_{r_{\rm ini}}^{r_{\rm fin}} \frac{\mathrm{d}r}{v_{r,{\rm d}}}, \nonumber \\
&&\approx 3.3\times 10^5 (1+\epsilon)^2 
\left[ \left( \frac{r_{\rm fin}}{1 \, {\rm au}} \right)^{-\frac{1}{2}} - \left( \frac{r_{\rm ini}}{1 \, {\rm au}} \right)^{-\frac{1}{2}} \right] {\rm yr},
\end{eqnarray}
where $r_{\rm ini}$ ($r_{\rm fin}$) is the initial (final) location of chondrules for their radial migration.
We set $r_{\rm fin}$ as $0.3$ au which is smaller than the orbital distance of Mercury.
The crossing time starting from $1$ au and from $100$ au are $\sim 2.6 \times 10^{5}$ yr and $6 \times 10^{5}$ yr, respectively.
Despite the two orders of magnitude difference between the initial locations, the crossing times differ only by about a factor of two.
\edit1{
Note that \citet{2012A&A...537A..61L} derived the migration time of a single dust particle considering the dependence on the properties of the gas disk (i.e., the dependence on the radial profile of surface density and temperature).
We estimate the migration time in a simplified way by assuming the MMSN as the solar nebula, because our target is the possibility of whether the dust particle moves within the survival time of chondrules.
Even if we ignore the variety of the radial profile of the gas disk, the qualitative results will not significantly change.
}

In this paper, we assume that chondrules are initially formed in the inner region of the solar nebula, and this is because the initial orbital radius of chondrules is unknown.
Nevertheless, as discussed in this section, the dependency of the crossing timescale of drifting chondrules on the initial orbital radius is relatively weak.
Chondrules may quickly grow into larger bodies due to the icy grain rim in the outer part of the solar nebula.
While this inappropriate growth might be a disadvantage, it is also a slight advantage in increasing their lifetime at the outer solar nebula.

\subsection{Plausible disk structure for chondrules surviving}
\label{sec:plaus-disk-struct}
\subsubsection{Comparison with observed features}
\label{sec:comp-with-observ}

In Section~\ref{sec:result}, we constrain the plausible combination of parameter range of $Z$, $\eta$, and $\alpha$ for the solar nebula by the evidence from chondrules (Figure~\ref{fig:eta_crit}).
We now compare the plausible condition for $Z$, $\eta$, and $\alpha$ for the solar nebula with the observational characteristics (or trends) for these values in the extrasolar systems.

First, we take $Z$ constrain from the disk observations.
The disk mass (particularly the mass of the gas component of the disk) is difficult to estimate.
Although it would be also difficult to refer to any statistical trends for $Z$ by observation, some $Z$ in the extrasolar disk is available.
In TW Hya disk, which is one of the most preferred disks for observations, $Z$ can be estimated from direct observations of dust and gas \citep{2017NatAs...1E.130Z}.
\citet{2020ApJ...893..125U} pointed out that $Z$ of the TW Hya disk can be large as $\lesssim 0.4$, taking into account the uncertainties in modeling the radiative transfer.
Here we consider $Z \lesssim 0.4$ as a credible constraint for the solar nebula.
\edit1{
Indeed, the TW Hya disk is a quite old and gas-depleted disk which is a suitable analog for the solar nebula at the time of chondrule formation.
}

Next, we constrain $\eta$ from observational features of the protoplanetary disks.
When we assume $\Sigma_{\rm g} \propto r^{-p}$ and $T \propto r^{-q}$ \edit2{(the same as Appendix~\ref{appendixA})}, we can rewrite Equation~(\ref{eq:eta_def}) as
\edit1{
\begin{eqnarray}
 \eta = \frac{1}{2}\left( \frac{3}{2}+p+\frac{q}{2} \right)\left(\frac{c_{\rm s}}{v_{\rm K}}\right)^{2}.
  \label{eq:eta_general}
\end{eqnarray}
}
The numerical coefficient of
\edit1{
$3/2+p+q/2$
}
relates to the radial slope of surface density and temperature of the protoplanetary disks.
The radial slope of surface density ($p$) is constrained by imaging with high spatial resolution.
The radial slope of temperature ($q$) is poorly constrained, and we here assume the optically thin disk ($q=1/2$) as usually assumed in observational modeling.
The dependency of $\eta \propto (c_{\rm s}/v_{\rm K})^{2} \propto (T/M_{\ast}) $ at the fixed $r$ comes from the fact that $\eta$ is the ratio between the radial pressure gradient force and the gravity from the host star.
Both $T$ and $M_{\ast}$ can be estimated by observations.

The radial slope of surface density for observed protoplanetary disks appears to be more gradual than the MMSN\footnote{Note that this is an estimate from the distribution of dust continuum emission (not from the gas distribution) in the radial direction.}.
Moreover, this tendency has been reported to be the case in several star-forming regions, Ophiuchus \citep{2009ApJ...700.1502A,2010ApJ...723.1241A}, Taurus-Auriga \citep{2019ApJ...882...49L}, and Lupus \citep{2017A&A...606A..88T}.
Even in the extreme case of $p=0$, however, the numerical coefficient of Equation~(\ref{eq:eta_general}) is of an order of unity.
Thus, $\eta$ can not be significantly changed by the radial slope of the disk.
Note that there is an exceptional disk structure having a reversal of the radial slope from the nominal protoplanetary disks (i.e., $p < 0$).
Although the reversal of the radial slope has been observed in fact\citep[e.g.,][]{2017A&A...606A..88T}, it is considered that such the reversal profile results from the disk dispersal at the late stage of the protoplanetary disk evolution.
Therefore here we consider that $p \geq 0$.

The disk temperature and the mass of the central star (i.e., the mass of the protosun) also can not be dramatically changed from the MMSN model.
In the outer region of the disk, the assumption of an optically thin disk (i.e., the assumption for the MMSN model), which is the standard assumption used in observational modeling, is considered to be sufficient \citep[e.g.,][]{2017ApJ...851...83C,2017A&A...606A..88T,2019MNRAS.482..698C}.
The temperature used to describe the observations and that of the MMSN is approximately the same in that region.
In the inner regions of the disk, on the other hand, the assumption of the optically thin disk breaks down.
The temperature estimated under the standard assumption should be different from the actual temperature profile of the observed disk. 
Nevertheless, the difference between the temperature of the actual disk and that of the MMSN model is likely to be a factor of several at most.

Overall, the observations support $\eta \approx 10^{-3}$, which is slightly smaller than that of the MMSN model: 
The observations suggest a disk with a shallow radial slope, with a central stellar mass and temperature profile similar to the MMSN model. 
In other words, chondrules could survive in such a disk.
Indeed, for disks with $p \lesssim 2/5$, $\eta$ becomes smaller than $\eta_{\rm c, max}$, so that there are viable parameters for the chondrule surviving.

Lastly, we also constraint on $\alpha$.
Although examples of observational constraints on turbulent strength are few and large uncertainties remain, there are emerging reports that $\alpha \lesssim 10^{-3}$ which is weaker than the classical value of $10^{-2}$ \citep[][and references therein]{2020ARA&A..58..483A}.
It should be pointed out that the recent disk evolution model that takes into account non-ideal MHD effects also supports a weak turbulence \citep[e.g.,][]{Bai2017a}.
The weak turbulence acts in the direction of lowering $Z_{\rm max}$, which can narrow the parameter range in which chondrules can survive in the $\eta$-$Z$ plane.
The combinations of $Z$ and $\eta$ that are both consistent with observations and chondrule-survivable are well restricted.

To summarize the features of observations so far, combinations such as $Z \sim 0.1$, $\eta \approx 10^{-3}$, and $\alpha \lesssim 10^{-3}$ may be natural features of protoplanetary disks.
This can be a sweet spot for chondrules to survive as shown in Figure~\ref{fig:eta_crit}.
From the perspective of chondrule surviving, we might say that the solar nebula had a structure similar to that of a typical protoplanetary disk.

\subsubsection{Variation of the gas disk mass}
\label{sec:variation-gas-disk}

We discuss \edit1{an} alternative scenario for the survival of the chondrules by changing the gas density of the solar nebula.
This is because there is a large uncertainty in the mass of the gas component of the protoplanetary disks.
Although we mainly treat the Stokes number as a function of the size of the dust particles in this paper, we obtain the relation of $t_{\rm drift} \propto \rho_{\rm g} T^{1/2}$ by Equations~(\ref{eq:t_stop_def}) and (\ref{eq:drift_time}).
We rewrite the Stokes number to reveal the dependency on the gas surface density as
\begin{eqnarray}
 \tau_{\rm stop} && = 2.71 \times 10^{-4} f^{-1} \left( \frac{r}{1\, {\rm au}} \right)^{3/2}
 \left( \frac{a}{1\, {\rm mm}} \right) \nonumber \\
 && \max \left[1, \, 3.19\times 10^{-2} f\left(\frac{a}{1 \, {\rm mm}}\right)\left(\frac{r}{1\, {\rm au}}\right)^{11/4} \right],
\label{eq:stokes_gas_dep}
\end{eqnarray}
where $f=\Sigma_{\rm g}/\Sigma_{\rm g, MMSN}$ is the factor to examine the disk mass (the same as we define in Section~\ref{sec:stre-inst}).
Chondrules with $1 \ {\rm mm}$ sizes are still in the Epstein regime when $f \lesssim 10^2$.
When $f \gtrsim 10$, however, the Stokes number is small enough for chondrules surviving of $10^{6}$ yr (see Equation~(\ref{eq:drift_time})).
Thus, $10$ times more massive disk than the MMSN is one of the promising \edit1{environments} for the chondrule survivability.

Oppositely, the self-gravitational instability of the solar nebula can not be negligible in such massive disk cases.
We estimate Toomre's $Q$ parameter \citep{Toomre1964a} for the solar nebula as\footnote{Note that the temperature profile is fixed here. Massive disks can have a situation where the assumption of an optically thin disk is compromised, and the temperature profile is varied.}
\begin{eqnarray}
Q = \frac{c_{\rm s}\Omega_{\rm K}}{\pi Gf \Sigma_{\rm g, MMSN}} \approx 5.6 \times 10 f^{-1} \left( \frac{r}{1\, {\rm au}} \right)^{-1/4}.
\label{eq:Qvalue}
\end{eqnarray}
For the massive disk with $f \gtrsim 10$, the $Q$ parameter becomes an order of unity.
Therefore the disk is in a marginally unstable state and can generate spiral arms.
These spiral arms might accumulate chondrules and be important for the size evolution of chondrules.

\edit1{
We also wish to discuss a possible scenario for the survivability of CAIs and chondrules.
Regarding the mass of the solar nebula, \citet{2011A&A...526L...8J} suggests that an initially massive solar nebula may explain two characteristics of CAIs; one is their narrow formation age and the other is their long lifetime.
Their long lifetime is achieved by dust trapping in the local region of the solar nebula.
On the contrary, it is necessary for chondrules to survive as free-floating particles for a long enough period without forming planetesimals (see Section~\ref{sec:dust-trap}).
We have proposed the preferred disk parameters in section~\ref{sec:comp-with-observ}, i.e., a disk that sufficiently slows radial drift of chondrule-sized particles.
Our preferred parameters may consistently combine with a CAI survival mechanism using dust trapping.
While chondrules survive by the slow radial drift, CAIs survive by trapping and subsequent diffusion in the disk.
The difference in their survival pathways can be \edit2{achieved} by the size sorting effects of radial drift (see also Section~\ref{sec:depend-radi-dust}).
In this case, both CAIs and chondrules can be maintained for the required period avoiding planetesimal formation.
However, a further study is necessary to find out whether the trapped CAIs can diffuse without forming planetesimals.
Otherwise, a planetesimal composed only of CAIs might form, which we have never seen.
}

\subsection{Planetesimal formation as subsequent evolution for chondrules' surviving}
\label{sec:plan-form-as}

Finally, we discuss possible pathways for the chondritic parent body.
\edit1{
Note that the formation processes of undifferentiated bodies is the target in this discussion, but that of differentiated bodies is beyond our scope.
}
After chondrules survive as chondrules for several million years in the solar nebula, in the end, chondrules should grow into larger bodies as planetesimals.
Chondrules provide the perspective that the appropriate pathways for the formation of the parent bodies of primitive chondrites.
Here are three planetesimal formation scenarios \citep[e.g.,][]{2014prpl.conf..547J}; (1) the SI and subsequent self-gravitational collapse of dense particle clumps, (2) the classical gravitational instability of the thin dust layer at the disk midplane, and (3) the collisional growth of dust particles due to the direct sticking.
The \edit1{typical} growth timescales of each mechanism are $\sim 10^{3}-10^{4}$ yr at most, which are far shorter than the survival time of chondrules.
These too-short growth timescales indicate that the planetesimal formation only occurs at a final phase of chondrules' life.
Consequently, the driving conditions for the formation of chondrites' parent bodies are established at a few millions years after the formation of the CAIs in the solar nebula.

We propose a scenario that the evolution of the solar nebula induces planetesimal formation.
One possible mechanism is the increase of the dust-to-gas mass ratio due to the global gas dispersal in the late stage of the disk evolution.
Such a high dust-to-gas mass ratio may drive the SI over a large region of the solar nebula.
Another mechanism is the late formation of disk substructures due to disk evolution.
Since the disk substructures can accumulate dust particles, these substructures are considered to lead to the planetesimal formation (see also Section~\ref{sec:dust-trap}).
Recent numerical simulations of protoplanetary disks support weak turbulent disks with magnetically-driven disk winds rather than fully turbulent disks (see Section~\ref{sec:plaus-disk-struct} and references therein).
These novel disk models suggest that the magnetically-driven disk winds can form the substructures later in disk life \citep{Suzuki2016a,Suriano2018a,Taki_2021}.
Considering recent disk models, planetesimal formation suitable for chondrule survival may occur spontaneously in disk evolution with magnetically-driven disk winds.
It is essential to compare this modern picture of disk evolution with the material evidence of the solar system, and further study is expected.

\section{Conclusions}
\label{sec:conclusions}

We investigated the \edit1{properties} of the solar nebula which is plausible for the long lifetime of chondrules.
We investigated the radial profile of the solar nebula which is plausible for the long lifetime of chondrules.
Isotopic and petrological evidence from primitive meteorites suggests that the chondrules survived at least $10^{6}$ yr in the solar nebula.
However, there are two main problems for such the long lifetime of chondrules; the radial drift of small grains and the planetesimal formation.
The former sweeps chondrules off from the solar nebula, and the latter prevents chondrules from maintaining their size and surviving \edit1{as free-floating particles} in the solar nebula for a long time.
We considered the survival condition for chondrules against their radial drift: The radial drift timescale should be longer than the survival time ($t_{\rm drift} \geq t_{\rm surv}$).
To avoid the planetesimal formation via the streaming instability (SI) in the disk midplane, we restrict the disk metallicity, $Z$, and turbulent strength, $\alpha$.

We found that the minimum mass solar nebula (MMSN)-like disk, which has the same radial profile of the gas surface density and temperature as the MMSM model, fails to reproduce the long lifetime of chondrules.
To increase the radial drift timescale, a higher disk metallicity and/or stronger sedimentation is required.
\edit1{Nevertheless}, the SI condition, $\epsilon \geq 1$, is always achieved in the MMSN-like disk when $t_{\rm drift} \geq t_{\rm surv}$.
We claimed that the plausible condition of the solar nebula is qualitatively consistent with the recent observational suggestions for protoplanetary disks, i.e., the higher $Z$ and/or flatter radial profile with the weaker turbulence than the MMSN model.
For instance, $Z \approx 0.1$ and $\eta \lesssim 10^{-3}$ with $\alpha \sim 10^{-6}$ are favorable to prevent chondrules from the radial drift and further growth.

We emphasize that the timing of planetesimal formation is important.
The early planetesimal formation is unlikely for the chondrules in terms of the thermal history of the parent bodies of chondritic meteorites.
It indicates that the conditions for the planetesimal formation are satisfied in the latter stages of the solar nebula evolution.
Based on the chondrule evidence, we can propose the following scenario for the history of the early solar system:
(1) In the early stage of the solar nebula evolution ($\lesssim 10^{6}$ yr), the radial profile of the solar nebula is suitable for the long lifetime of chondrules (Section~\ref{sec:plaus-disk-struct}).
(2) In the late stage of the solar nebula evolution ($\gtrsim 10^{6}$ yr), the disk evolution succeeds in achieving the conditions for the formation of planetesimals.
Planetesimal formation associated with the inside-out dispersal of the gaseous disk is one possible scenario for the formation mechanism of planetesimals consistent with the chondrule evidence (Section~\ref{sec:plan-form-as}).

\acknowledgments
We thank Taishi Nakamoto, Satoshi Okuzumi, and Eiichiro Kokubo for helpful suggestions on an early version of this work.
TT is also grateful to Sota Arakawa for a fruitful discussion on the nature of chondrules.
This work was supported by JSPS KAKENHI Grant Number 21K13983, 22H05150.

\appendix
\section{Effect of the gas accretion velocity}
\label{appendixA}
\edit2{
We estimate the gas radial velocity due to the viscous accretion as
\begin{eqnarray}
 u_{\nu} \sim -\frac{\nu}{r} = -\alpha \left(\frac{h_{\rm g}}{r}\right)^{2} v_{\rm K}.
\end{eqnarray}
Assuming that the radial profiles of gas surface density and temperature are $\Sigma_{\rm g} \propto r^{-p}$ and $T \propto r^{-q}$,
we can rewrite Equation~(\ref{eq:NSH}) as
\begin{eqnarray}
 v_{{\rm d}, r } \sim  
  - \frac{2\tau_{\rm stop}}{(1+\epsilon)^2 + \tau_{\rm stop}^2} \eta v_{\rm K}
\left[
   1 + \frac{2(1+\epsilon)}{2p+q+3}\left(\frac{\alpha}{\tau_{\rm stop}}\right)
\right].
\end{eqnarray}
The second term in the bracket corresponds to the effects of the gas accretion.
The coefficient in front of $\alpha/\tau_{\rm stop}$ is usually an order of unity.
Therefore, the effects of the gas accretion on the dust radial velocity switch around $\alpha \approx \tau_{\rm stop}$.
}

\edit2{
As we show in Section \ref{sec:result}, our main interest is the situation where $\alpha \lesssim \tau_{\rm stop}$. 
Our favorable conditions for surviving chondrules in the disk are $\alpha \ll \tau_{\rm stop}$ (see Section \ref{sec:results-fiduc-model} and Figure~\ref{fig:contour}).
In addition, the dust radial velocity varies only a few times, even where $\alpha \approx \tau_{\rm stop}$.
This is the reason why we ignore the effects of gas accretion on the dust radial velocity.
}




\bibliography{ref}





\end{document}